\documentclass[a4paper]{spie}

\usepackage{amsmath,amsfonts,amssymb}
\usepackage{graphicx}
\usepackage[colorlinks=true, allcolors=blue]{hyperref}
\usepackage{siunitx}

\title{Towards a test of quantum gravity with a levitated nanodiamond containing a spin}

\author[a]{Benjamin D. Wood}
\author[a]{Gavin W. Morley}
\affil[a]{Department of Physics, University of Warwick, Coventry, CV4 7AL, United Kingdom}

\authorinfo{Further author information:\\B.D.W.: E-mail: ben.d.wood@warwick.ac.uk\\G.W.M.: E-mail: gavin.morley@warwick.ac.uk}

\begin{document} 
\maketitle

\begin{abstract}
There is significant interest in potential experimental tests of macroscopic quantum effects, both to test potential modifications to quantum theory and to probe the quantum nature of gravity. A proposed platform with which to generate the required macroscopic quantum spatial superposition is a nanodiamond containing a negatively charged nitrogen vacancy (${\text{NV}}^{-}$) centre. In this review, methods to fabricate nanodiamonds containing ${\text{NV}}^{-}$ suitable for these quantum applications are discussed. The proposed probes of the macroscopic limits of quantum theory are presented along with the spin physics of the ${\text{NV}}^{-}$ centre relevant to those tests.
\end{abstract}


\section{INTRODUCTION}
\label{sec:intro}  

Levitated nanodiamond experiments have been proposed as a platform to probe the macroscopic limits of quantum theory \cite{scala_2013, wan_2016b, wan_2016a, pedernales_2020, wood_2021, yin_2013} and the quantum nature of gravity \cite{albrecht_2014, bose_2017, marletto_2017}. These experiments utilise the electronic spin of the negatively charged nitrogen vacancy (${\text{NV}}^{-}$) defect in diamond. Whilst the levitated nanodiamond experiments suggested exhibit desirable features as a test of macroscopic superposition, there are still a number of challenges to be overcome before they are practically achievable. These include the short ${\text{NV}}^{-}$ spin coherence (${T}_{2}$) time in nanodiamonds compared to that of bulk diamond.

This review will provide a background to the synthesis of diamond, the optical and spin physics of the ${\text{NV}}^{-}$ centre, and the specific methods of nanodiamond fabrication in Sec. \ref{sec:DS}, \ref{sec:NV}, and \ref{sec:NDCNV} respectively. Next, the proposed experimental tests of the macroscopic limits of quantum physics and proposed probes of the quantum nature of gravity are discussed in Sec. \ref{sec:QL}. Finally, the different methods of trapping nanodiamond in order to perform the levitated nanodiamond experiments are discussed in Sec. \ref{sec:NDL}.

\section{DIAMOND SYNTHESIS}
\label{sec:DS}

Using synthetically grown diamond is advantageous for research purposes as the impurity content and crystal morphology can be controlled. This is in contrast to naturally formed diamond where, by its nature, the growth conditions cannot be controlled. There are two primary methods of bulk diamond synthesis, high pressure high temperature (HPHT) synthesis and chemical vapour deposition (CVD) synthesis. Both techniques allow the nitrogen content of the grown diamond to be controlled. This is important when trying to control the final ${\text{NV}}^{-}$ defect concentration in a diamond sample.

\subsection{High Pressure High Temperature}
\label{sec:HPHT}

One method to synthesise diamond is to place carbon in a high pressure high temperature (HPHT) environment with a metal solvent \cite{bundy_1955}. The high temperatures and pressures are required as at room temperature and pressure the thermodynamically stable form of carbon is graphite, ${sp}^{2}$ bonded carbon. Here each carbon atom bonds to its three equidistant nearest neighbours in a plane. For higher pressures diamond, ${sp}^{3}$ bonded carbon, is the thermodynamically stable form. Each carbon bonds to four equidistant nearest neighbours in a tetrahedral structure \cite{bundy_1996}. Modern HPHT reactors typically operate in the parameter space of $1600 - \SI{1900}{\kelvin}$ and $5 - \SI{6}{\giga\pascal}$ \cite{ashfold_2020, stromann_2006, schmetzer_2010}, however synthesis at higher temperatures and pressures is possible \cite{palyanov_2015}.

Nitrogen is the dominant impurity in HPHT diamond, forming mostly single substitutional nitrogen in typical HPHT reactors, however at extremely high temperatures the substitutional nitrogen can aggregate \cite{chrenko_1977}. The typical nitrogen concentrations incorporated into HPHT diamonds are approximately $200-300 ~\text{ppm}$ which can be further increased by adding nitrogen containing compounds to the growth mixture.

As well as adding compounds to increase the nitrogen incorporation into the final diamond material when trying to generate a high nitrogen content, it can be desirable to grow higher purity diamond where it is required to try and minimise the nitrogen content. This is achieved by adding a nitrogen `getter' material into the carbon source material. This is typically Al or Ti metal as they both form stable nitrides \cite{strong_1971}.

\subsection{Chemical Vapour Deposition}\label{sec:CVD}

An alternative diamond synthesis technique to HPHT is CVD. CVD is a technique to grow diamond material in the region of phase space where graphite is the thermodynamically stable form of carbon. Whilst this is not possible in the context of equilibrium thermodynamics, chemical kinetics allow ${sp}^{3}$ carbon to be preferentially grown over ${sp}^{2}$ carbon \cite{angus_1968}.

The CVD process utilises mostly H and C containing source gases along with a growth seed. The source gases are dissociated into their radical components by transferring energy to the gas mixture. This can be done either by a hot filament running close to the growth seed \cite{haubner_1993}, or by placing the seed within a microwave resonator cavity \cite{tallaire_2013}. The microwave radiation or hot filament ignites a plasma of gas radicals at low pressure, $< 1$ atm. \cite{ashfold_2020}. The typical CVD growth process involves hydrogen radicals etching carbon-hydrogen bonds on the surface of the diamond growth, then the majority of the time a hydrogen radical will reform the C-H bond. However, sometimes a carbon radical can be captured by the dangling bond, forming a C-C bond. This bonding can initially be ${sp}^{2}$ or ${sp}^{3}$, but ${sp}^{2}$ bonds are etched away by H radicals faster than the etch rate of ${sp}^{3}$ bonded carbon \cite{ashfold_2020, tallaire_2013}. This generates a net rate of growth of diamond material. 

By adding N containing gases to the H and C containing source gases, with precise control, CVD-grown single crystal diamonds with single substitutional nitrogen content ranging from $< 0.5 ~\text{ppb}$ to $> 1 ~\text{ppm}$ are available commercially. The high purity achievable by the CVD technique has made CVD-produced diamonds a useful platform for quantum applications, in particular quantum sensing \cite{achard_2020}.

\section{Nitrogen Vacancy Centre}\label{sec:NV}

The nitrogen vacancy (NV) centre is a defect in the carbon lattice of diamond. It consists of a substitutional nitrogen atom on a lattice site in diamond that has a vacant lattice site as one of its nearest neighbours \cite{rondin_2014}. The electronic structure of the neutral defect (${\text{NV}}^{0}$) is made up of three electrons contributed from dangling carbon bonds around the vacancy, and two contributed from the valence electrons in nitrogen after forming three bonds with adjacent carbons \cite{larsson_2008}. ${\text{NV}}^{0}$ can accept an electron from a donor species in the carbon lattice to form the negatively charged ${\text{NV}}^{-}$ variant of the nitrogen vacancy centre \cite{ashfold_2020}. The ${\text{NV}}^{-}$ centre is the subject of growing interest in the fields of magnetometry \cite{rondin_2014}, quantum networking and computing \cite{awschalom_2018}, and probes of fundamental physics \cite{scala_2013, wan_2016b, wan_2016a, wood_2021}. 

The ${\text{NV}}^{-}$ centre can be found as a grown-in defect from the chemical vapour deposition (CVD) diamond growth process or it can be created in existing diamond material \cite{doherty_2013}. The details of these methods of NV creation are covered in Sec. \ref{sec:NVC}.

\begin{figure}
\begin{center}
\includegraphics[width=0.33\textwidth]{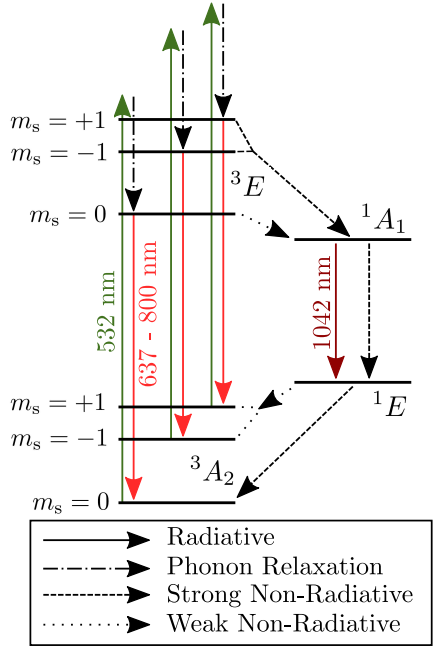}
\end{center}
\caption{Energy level diagram of the ${\text{NV}}^{-}$ centre. The degenerate ${m}_{\text{s}} = \pm 1$ levels are zero-field split from ${m}_{\text{s}} = 0$. Degeneracy is broken by application of an external magnetic field. Under $\SI{532}{\nano\metre}$ excitation the relative strengths of non-radiative transitions mean that initially the strength of $637 - 800 ~\si{\nano\metre}$ fluorescence can be used to readout the spin state, and that after continued excitation the spin state is polarised to ${m}_{\text{s}} = 0$.}
\label{fig:nvlevels}
\end{figure}

The ${\text{NV}}^{-}$ centre is an electronic spin-1 system \cite{doherty_2013}, with energy levels shown in Fig. \ref{fig:nvlevels}. The ground state spin triplet, ${}^{3}{\text{A}}_{2}$, has a zero field splitting of $\SI{2.87}{\giga\hertz}$ \cite{doherty_2013}.

\subsection{Optically Induced Spin Polarisation}\label{sec:OISP}

The ${\text{NV}}^{-}$ centre exhibits optically induced spin polarisation; the process is depicted in Fig. \ref{fig:nvlevels}. To optically polarise, first excite the ${\text{NV}}^{-}$  with $\SI{532}{\nano\metre}$ laser light. This drives the system from the ground ${}^{3}{\text{A}}_{2}$ state to the phonon band of the ${}^{3}\text{E}$ excited state. This optical transition is spin preserving \cite{rondin_2014}. After the excitation, the electron will quickly undergo phonon relaxation and fall to the corresponding zero-phonon ${m}_{\text{s}} = \pm1$ or $0$ spin state of the ${}^{3}\text{E}$ level.

In the case that the ground state electron was initially in the state ${m}_{\text{s}} = 0$, the most likely channel of decay to the ground state is through a spin-preserving optical transition back to the ${m}_{\text{s}} = 0$ spin state of the ${}^{3}{\text{A}}_{2}$ level \cite{robledo_2011, doherty_2013}. This transition emits a photon into the $\SI{637}{\nano\metre}$ zero phonon line (ZPL) of the ${\text{NV}}^{-}$ centre, or into the phonon sideband of the $\SI{637}{\nano\metre}$ ZPL that extends to $\approx \SI{800}{\nano\metre}$ \cite{beha_2012}. In this case the system remains in the ${m}_{\text{s}} = 0$ spin state and will undergo cycles of photoluminescence when excited by $\SI{532}{\nano\metre}$ light.

In the case that the system was originally in the ${m}_{\text{s}} = +1$ or $-1$ state, the non-radiative inter-system crossing (ISC) decay channel from the ${m}_{\text{s}} = \pm 1$ excited states to the ${}^{1}{\text{A}}_{1}$ energy level is significantly more likely than the system undergoing ISC from the ${m}_{\text{s}} = 0$ excited state \cite{robledo_2011, doherty_2013}. From the ${}^{1}{\text{A}}_{1}$ energy level, the system undergoes a transition to the ${}^{1}\text{E}$ state. Finally the system undergoes another ISC transition to the ${m}_{\text{s}} = 0$ or ${m}_{\text{s}} = \pm 1$ ${}^{3}{\text{A}}_{2}$ ground state with approximately equal probability \cite{robledo_2011, doherty_2013}. Therefore under continued $\SI{532}{\nano\metre}$ excitation the ${\text{NV}}^{-}$ centre spin polarises to the ${m}_{\text{s}} = 0$ state.

Other decay channels are possible, as shown in Fig. \ref{fig:nvlevels}, therefore the exact degree of polarisation that can be achieved is variable, with values from $42\% - 96\%$ reported \cite{doherty_2013}.

\subsection{Microwave Control}\label{sec:MC}

To manipulate the spin state of the ${\text{NV}}^{-}$ centre, a static magnetic field is applied to the host diamond. This splits the otherwise degenerate ${m}_{\text{s}} = \pm1$ levels due to Zeeman splitting \cite{rondin_2014}. The ${m}_{\text{s}} = \pm1$ states move away from the zero magnetic field state at $\pm \SI{28.02}{\mega\hertz\per\milli\tesla}$ \cite{sangtawesin_2016}. An  applied microwave field, tuned to be resonant either with the ${m}_{\text{s}} = 0 \leftrightarrow +1$ or ${m}_{\text{s}} = 0 \leftrightarrow -1$ transition, will flip the spin state.

\subsection{Spin Dependant Fluorescence}\label{sec:SDF}

To readout the spin state of the ${\text{NV}}^{-}$ centre, optical detection is typically used. If the ${\text{NV}}^{-}$ centre spin state is in ${m}_{\text{s}} = 0$, as a $\SI{532}{\nano\metre}$ excitation is switched on the ${\text{NV}}^{-}$ centre will produce photoluminescence in the $\SI{637}{\nano\metre} - \SI{800}{\nano\metre}$ range. This is due to the ${m}_{\text{s}} = 0$ state cycling from excited state to ground state as described in Sec. \ref{sec:OISP}. If the ${\text{NV}}^{-}$ centre is in the ${m}_{\text{s}} = +1$ or $-1$ state and the $\SI{532}{\nano\metre}$ excitation is switched on, before the spin state optically polarises the $\SI{637}{\nano\metre} - \SI{800}{\nano\metre}$ photoluminescence produced will be of a lower intensity than for the ${m}_{\text{s}} = 0$ case. Also the ${}^{3}\text{E}$, ${m}_{\text{s}} = 0$ optical decay lifetime is $\SI{23}{\nano\second}$ in nanodiamond \cite{neumann_2009} and $\SI{12}{\nano\second}$ in bulk \cite{doherty_2013}, whereas the limiting decay lifetime of the ${}^{1}{\text{A}}_{1}$, ${}^{1}\text{E}$ pair of levels is $\approx \SI{300}{\nano\second}$ \cite{rondin_2014}. Therefore, the ${m}_{\text{s}} = \pm 1$ state gets `stuck' in the long lifetime ${}^{1}{\text{A}}_{1}$, ${}^{1}\text{E}$ decay channel, further decreasing the photoluminescence signal.

By exciting the ${\text{NV}}^{-}$ centre with $\SI{532}{\nano\metre}$ laser light, frequency sweeping a microwave signal to manipulate the ${\text{NV}}^{-}$ spin state and detecting $\SI{637}{\nano\metre} - \SI{800}{\nano\metre}$ emission the techniques of optically-induced spin polarisation, microwave control, and spin dependant fluorescence are used to facilitate optically detected magnetic resonance (ODMR). An example ODMR spectrum of the ${\text{NV}}^{-}$ centre is shown in Fig. 2 of Ref. \citenum{rondin_2014}.

\subsection{Power Saturation}\label{sec:PS}

The photoluminescent emission of ${\text{NV}}^{-}$ centres exhibits saturation as the excitation intensity increases, see Fig. 5 of Ref. \citenum{jelezko_2006}. There is a maximum number of excitation and decay cycles that can occur per second, due to the finite state lifetimes, limiting the maximum emission count rate no matter how much power the excitation laser delivers. The emitted photon count rate $I$ as a function of excitation power $P$ is fit with eqn. \ref{eq:Psat} \cite{plakhotnik_2018}:
\begin{equation}
I = {I}_{\infty}\frac{P}{{P}_{\text{Sat}}+P}\label{eq:Psat}
\end{equation}
where ${I}_{\infty}$ and ${P}_{\text{Sat}}$ are the photon count rate for infinite excitation power and the excitation saturation power respectively.

\subsection{Photon Correlation}\label{sec:PC}

To determine whether observed signals originate from a single ${\text{NV}}^{-}$ centre a Hanbury Brown-Twiss (HBT) measurement is used \cite{hanbury_1956}\cite{stephen_2019}. The emitted photons are split into two channels, each of which are detected as clicks in single photon detectors, and the time of detection logged. By histogramming the time delay between a click in one channel and a click in the other the degree of correlation between photons with a given delay time can be established. The second order correlation function, ${g}^{(2)}(\tau)$, for this process is given by eqn. \ref{eq:g2} \cite{berthel_2015}:
\begin{equation}
{g}^{(2)}(\tau) = \frac{\langle I(t+\tau)I(t)\rangle}{{\langle I(t)\rangle}^{2}}\label{eq:g2}
\end{equation}
where $\tau$ is the time delay between two clicks, and $I(t)$ is the electromagnetic energy absorbed by a detector at time $t$. $\langle \dots \rangle$ indicates that the process is averaged over time. ${g}^{(2)}(\tau) > 1$ indicates a positive correlation between photon detection events separated by $\tau$, ${g}^{(2)}(\tau) < 1$ indicates negative correlation, see Fig. 6 of Ref. \citenum{jelezko_2006}, and ${g}^{(2)}(0) = 0$ indicates that two photons were never detected simultaneously, one at each detector \cite{berthel_2015}.  A single ${\text{NV}}^{-}$ centre can only emit one photon at a time, so ideally ${g}^{(2)}(0) = 0$. An ideal pair of ${\text{NV}}^{-}$ centres exhibit ${g}^{(2)}(0) = 0.5$, therefore allowing for noise a single centre is identified if ${g}^{(2)}(0) < 0.5$ \cite{stephen_2019}.

\subsection{Spin Coherence Time}\label{sec:SCT}

The electron spin coherence time, ${T}_{2}$, of the ${\text{NV}}^{-}$ centre is a measure of the time that the spin can store a quantum state without it decohering due to other spins in its local environment \cite{wang_2005}. The ${T}_{2}$ time can be measured directly by initialising the ${\text{NV}}^{-}$ into the ${m}_{\text{s}} = 0$ state, applying a Hahn echo microwave pulse sequence \cite{hahn_1950}, and using spin-dependant fluorescence to readout the degree of coherence of the ${\text{NV}}^{-}$ spin state.

The ${T}_{2}$ time places a limit on the macroscopicity of spatial superposition that can be created using the proposed methods of Sec. \ref{sec:ET}. The ${T}_{2}$ time can be extended by, instead of using a simple Hahn echo pulse sequence, using a dynamical decoupling pulse sequence that better decouples the ${\text{NV}}^{-}$ spin from the local spin environment. For example the CMPG or XY8 dynamical decoupling sequences \cite{stephen_2019, bar-gill_2013, wang_2012} act to cancel the phase contributions of environmental spins. To maximise the ${T}_{2}$ time, pulse sequences containing on the order of ${10}^{3} - {10}^{4}$ $\pi$ pulses are required \cite{bar-gill_2013}. Figure 11 of Ref. \citenum{stephen_2019} shows an number of example measurements of the ${T}_{2}$ time of a single ${\text{NV}}^{-}$ electronic spin using the Hahn echo pulse sequence.   

One source of decoherence is from the spin-$\frac{1}{2}$ ${}^{13}\text{C}$ nuclei \cite{childress_2006} that occur naturally in diamond with $1.1\%$ abundance \cite{stanwix_2010}.  Other factors can also contribute to limiting ${T}_{2}$ time, such as the concentration of substitutional nitrogen defects and the degree to which the external static magnetic field used to induce Zeeman splitting, see Sec. \ref{sec:MC}, is aligned with the ${\text{NV}}^{-}$ axis \cite{maze_2008}. If the field is off axis the ${m}_{\text{s}} = 0, \pm 1$ states are no longer the eigenstates of the ${\text{NV}}^{-}$ spin, causing additional decoherence. 

The longest ${\text{NV}}^{-}$ electron spin ${T}_{2}$ time measured is $\SI{1.58}{\second}$ using a $< 5$ ppb nitrogen concentration, CVD grown, bulk diamond cooled to $\SI{3.7}{\kelvin}$ \cite{abobeih_2018}. The observed ${T}_{2}$ times in nanodiamonds are significantly shorter than those for bulk diamond. A proposed explanation for these differences is that impurities and dangling bonds at the diamond surface can modify the charge transport properties at the surface, causing an increase in the number of ${\text{NV}}^{-}$ centres that are charge switched to their neutrally charged counterpart \cite{rondin_2010}. Charge transport can also induce electric field noise at the ${\text{NV}}^{-}$ centre, along with surface spin flips that induce magnetic field noise. Both of these noise sources reduce the nanodiamond ${T}_{2}$ time.

The longest ${T}_{2}$ time in nanodiamonds, using CVD growth and lithography techniques to manufacture isotopically pure ${}^{12}\text{C}$ diamond cylinders of diameter $\approx 500 ~\si{\nano\metre}$ and length $\approx 1 ~\si{\micro\metre}$, is ${T}_{2} = \SI{708}{\micro\second}$ when using dynamical decoupling techniques \cite{andrich_2014}. For natural abundance ${}^{13}\text{C}$ nanodiamonds, the longest ${T}_{2}$ time reported is $\SI{462}{\micro\second}$ measured in nanodiamonds produced by CVD growth and milling \cite{wood_2021}. The nanodiamond containing the ${\text{NV}}^{-}$ with ${T}_{2} = \SI{462}{\micro\second}$ had diameter $\approx \SI{200}{\nano\metre}$. Both of these measurements were made at room temperature.

\section{Nanodiamond Containing ${\text{NV}}^{-}$}\label{sec:NDCNV}

\subsection{${\text{NV}}^{-}$ Creation}\label{sec:NVC}

The NV centre can be found as a grown-in defect from the chemical vapour deposition (CVD) diamond growth process \cite{johansson_2010}, or created as a product of particle irradiation and annealing \cite{edmonds_2021}, as a product of nitrogen ion implantation and annealing \cite{vandam_2019}, and by laser writing \cite{chen_2017}. 

In both the irradiation and ion implantation cases the principle of the process is the same. The methods ensure that substitutional nitrogen and vacancy sites are present in the diamond lattice, then an anneal at $> \SI{600}{\celsius}$ makes the vacancies mobile in the diamond lattice \cite{edmonds_2021, capelli_2018}. The vacancies can move until they are trapped by a substitutional nitrogen, forming an NV centre. The centres form in the ${\text{NV}}^{-}$ or ${\text{NV}}^{0}$ charge states \cite{edmonds_2021} and therefore trying to maximise the creation of ${\text{NV}}^{-}$ over ${\text{NV}}^{0}$ is important when the diamond is to be used for ${\text{NV}}^{-}$ applications. Despite vacancies being mobile for temperatures $> \SI{600}{\celsius}$ it has been observed that the annealing temperatures to maximise the number of NV centres produced are those $> \SI{800}{\celsius}$ \cite{botsoa_2011, orwa_2010}. 

In the irradiation case, vacancies are formed by high energy particle irradiation and the substitutional nitrogen content is provided by the grown-in nitrogen content of the diamond sample. The density of NV centre creation is limited by the concentration of grown-in substitutional nitrogen. In the ion implantation case, nitrogen ions are implanted into the diamond lattice \cite{chu_2014}. This process has the benefit of allowing the final substitutional nitrogen content to be controlled, and the kinetics of implanting nitrogen ions will also create vacancy centres in the same region. 

Laser writing of NV centres differs slightly from the processes discussed above. Vacancy centres are created in the lattice by using a focused, high energy, laser pulse. A global anneal in a furnace \cite{stephen_2019, chen_2017} or a pulse train of lower energy pulses follows the initial vacancy-creating pulse to anneal the area local to the created vacancy \cite{chen_2019}. Laser-written ${\text{NV}}^{-}$ centres can have ${T}_{2}$ times \cite{stephen_2019} and optical coherence \cite{chen_2017} as good as naturally occurring ${\text{NV}}^{-}$.

\subsection{Nanodiamond Fabrication}\label{sec:NF}

There are two primary methods of nanodiamond fabrication for quantum applications. The first method of nanodiamond production is to synthesise a macroscopic parent diamond sample, either by HPHT \cite{knowles_2014} or CVD \cite{frangeskou_2018} methods, and then to mechanically mill the sample into nano-scale particles \cite{gines_2018}. In this scheme the ${\text{NV}}^{-}$ concentration in the parent diamond sample and the final size of the nanodiamond particles sets the probability that a nanodiamond contains a certain number of ${\text{NV}}^{-}$ centres. Typically when creating nanodiamonds with the aim of measuring a long ${\text{NV}}^{-}$ ${T}_{2}$ time, the parameters are chosen such that the probability of a nanodiamond containing any ${\text{NV}}^{-}$ centres is low. This is because having pure diamond material is desired and measurements can be repeated until a nanodiamond containing an ${\text{NV}}^{-}$ is found experimentally.

The other method for nanodiamond fabrication is based on etching of a specifically grown parent sample. It can be used to create nanodiamonds with tailored shape and size \cite{andrich_2014}. One example is the use of electron beam lithography to pattern an etch mask onto a sample of CVD grown diamond membrane that includes a $\sim \SI{6}{\nano\metre}$ thick ${\text{NV}}^{-}$ defect layer in the centre of the vertical profile of the membrane. The prepared sample is then plasma etched. This removes the material in the unmasked regions. The mask pattern used results in cylindrical nanodiamond particles that are $\sim \SI{1}{\micro\metre}$ in depth and $\sim \SI{500}{\nano\metre}$ in diameter. 

Comparing the two methods, a disadvantage of the milling technique over etching techniques is that there is no control over the placement of the ${\text{NV}}^{-}$ within the nanodiamond or the final shape of the nanodiamond particle, however unlike etching, milling techniques allow creation of nanodiamonds from the full 3D volume of the bulk material at once. It is therefore easier to create large quantities of nanodiamonds with milling. In trapping applications where nanodiamonds are loaded by spraying \cite{hsu_2016}, the large number of nanodiamonds created in one milling run is beneficial as thousands of nanodiamonds are sprayed to trap one.

\section{Quantum Limits}\label{sec:QL}

Quantum mechanics is a hugely successful theory, however there are still elements that warrant further experimental investigation. For instance, quantum theory places no limit on the allowed macroscopicity of superpositions, however molecules containing around $2000$ atoms have demonstrated the most macroscopic superposition as yet observed \cite{haslinger_2013, eibenberger_2013, fein_2019}.

The lack of observation of macroscopic superposition could be explained by larger systems not being isolated from the environment well enough, however, there are also a number of proposed modifications to quantum mechanics. A family of modifications can be described as spontaneous collapse models, for example continuous spontaneous localisation (CSL), where terms added to the evolution can spontaneously collapse superposition states with a characteristic rate and length scale \cite{bassi_2013, frowis_2018}. By observing more macroscopic superpositions in the lab, bounds can be placed on the parameters of the proposed modifications to quantum theory. 

The degree to which an experimental observation bounds the time scale of spontaneous collapse, given the length scale of the experiment, can be used to define a measure of the macroscopicity of the superposition \cite{nimmrichter_2013}. It is by this measure that the $2000$ atom experiment referenced previously is the most macroscopic \cite{fein_2019}. 

Another outstanding issue with quantum theory is the difficulty of combining it with general relativity. To date, there is a lack of experimental observations that probe the quantum nature of gravity that can be used to help guide theoretical investigation. One such probe of the quantum nature of gravity would be an experiment that could determine whether the gravitational interaction can entangle two quantum states \cite{bose_2017, marletto_2017}.

\subsection{Experimental Tests}\label{sec:ET}

\begin{figure}
\begin{center}
\includegraphics[width=0.8\textwidth]{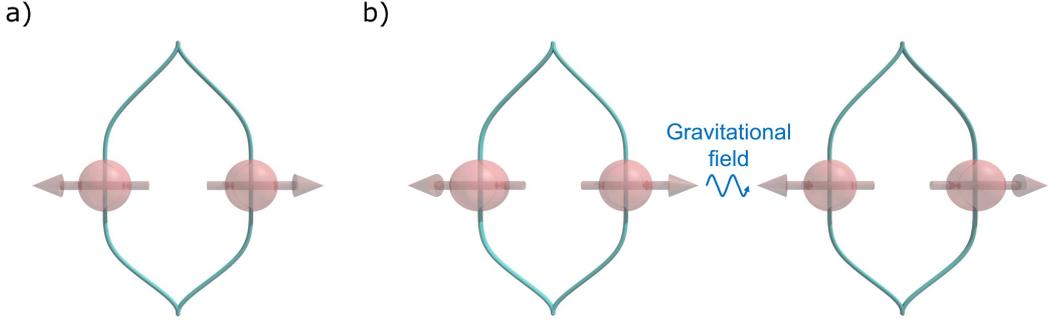}
\end{center}
\caption{Schematics of the proposed schemes to test the limits of quantum mechanics with nanodiamonds containing ${\text{NV}}^{-}$. (a) To test the macroscopic limits of superposition, a magnetic field gradient would act on an ${\text{NV}}^{-}$ spin superposition creating a superposition of forces, spatially splitting the superposition components into a spatial superposition of the nanodiamond \cite{scala_2013, wan_2016b, wan_2016a, pedernales_2020, wood_2021, yin_2013}. The spatial superposition would then be recombined such that at the end of the scheme the spin state could be readout optically. After multiple runs with different tilts of the experiment with respect to the vertical defined by gravity, a spin interference fringe pattern would evidence the presence of the spatial superposition \cite{scala_2013, wan_2016b, wood_2021}. (b) To probe the quantum nature of gravity, a pair of macroscopic spatial superpositions would be created simultaneously in close proximity to each other. If the distances involved are of the required magnitudes, the gravitational interaction between the two superposition components at the closest approach of the two nanodiamonds is dominant over the gravitational interaction between other pairs of superposition components \cite{bose_2017, marletto_2017}. In this case the two ${\text{NV}}^{-}$ spin states could be entangled if the gravitational interaction is quantum in nature \cite{bose_2017, marletto_2017, marshman_2020b}.}
\label{fig:q_tests}
\end{figure}

The use of levitated nanodiamond particles to probe the limits of the quantum superposition principle has been proposed \cite{scala_2013, wan_2016b, wan_2016a, pedernales_2020, wood_2021, yin_2013}. The nanodiamond particle containing an ${\text{NV}}^{-}$ centre is levitated and then the ${\text{NV}}^{-}$ spin is manipulated to create a ${m}_{\text{s}} = \pm1$ spin superposition state. 

A magnetic field gradient is applied to the nanodiamond containing the ${\text{NV}}^{-}$ centre. The gradient exerts a force on the ${\text{NV}}^{-}$ spin magnetic moment in opposite directions depending on the ${m}_{\text{s}} = \pm1$ state. This places the nanodiamond into a spatial superposition, as shown in Fig. \ref{fig:q_tests}(a), and allows the nanodiamond to be used as a matter-wave interferometer. This Stern-Gerlach interferometry has been demonstrated experimentally for a Bose-Einstein condensate \cite{keil_2021, margalit_2021}.

By applying the magnetic field gradient at a variable angle with respect to the horizontal \cite{scala_2013, wan_2016b, wood_2021} the nanodiamond superposition could be directly detected, due to the accumulated phase difference as the two nanodiamond positions in the superposition experience different gravitational potential energies. The phase difference would be measured as an interference pattern of the final ${\text{NV}}^{-}$ spin state once the two superposition components have been re-combined.

The proposals include schemes in which the nanodiamond is trapped, or in free fall. The trapped schemes are closer to being practically achieved, and could be used to observe matter-wave interference of a large mass ($\approx \SI{1e-17}{\kilo\gram}$) nanodiamond \cite{scala_2013}, however, the superposition distance is limited, limiting the macroscopicity. The free fall schemes can increase the achievable superposition distance such that the macroscopicity of those experiments would be among the largest proposed to date and bounds could be tightened on the parameters of the CSL modification to quantum mechanics \cite{wan_2016b, pedernales_2020, wood_2021}.

Alongside tests of macroscopic superposition it has also been proposed that nanodiamonds containing ${\text{NV}}^{-}$ centres could be used to verify the quantum nature of gravity by checking for gravitational entanglement effects \cite{bose_2017, marletto_2017 , marshman_2020b}. This type of experiment will require a pair of nanodiamonds to both be placed into highly macroscopic superpositions within gravitational interaction range of each other, as shown in Fig. \ref{fig:q_tests}(b). Beyond the experimental challenges of generating a single superposition with a large separation for a long time \cite{marshman_2021, wood_2021}, it is crucial to ensure entanglement could only have been generated by gravitational effects rather than other interactions, such as casimir forces \cite{chevalier_2020, toros_2020, toros_2021, van_de_Kamp_2020}.     

\section{Nanodiamond Levitation}\label{sec:NDL}
\subsection{Optical Traps}\label{sec:O}

Optical traps can levitate nanodiamonds by inducing an electric dipole in the trapped particle. The electric dipole is then confined at the focal point of a high intensity laser beam, with trapping powers typically of the order of $\SI{100}{\milli\watt}$ \cite{frangeskou_2018, neukirch_2013, neukirch_2015, hoang_2016b, pettit_2017}. It was observed that the total photoluminescence rate measured from the ${\text{NV}}^{-}$ centre, and the ODMR contrast, decreased as the trapping laser power increased \cite{hoang_2016b}. The high intensity laser could cause ionisation of the  ${\text{NV}}^{-}$ centre to ${\text{NV}}^{0}$, reducing the measured ${\text{NV}}^{-}$ photoluminescence \cite{hoang_2016b}. The internal temperature of the nanodiamond can increase due to the heating effect of the trapping laser. Minimising heating as a source of decoherence in proposed tests of quantum superposition is crucial and so optical traps are no longer the ubiquitously suggested trapping mechanism in experimental proposals using levitated nanodiamonds. 

\subsection{Paul Traps}\label{sec:PT}

Paul traps create a radio frequency oscillating electric field to confine a charged particle. There are a number of different electrode designs that satisfy this requirement, with those that have been used to trap nanodiamonds including a needle electrode design \cite{delord_2017b}, an end cap geometry \cite{conangla_2018}, linear trap \cite{kuhlicke_2014}, and a ring design \cite{delord_2017a, delord_2018, delord_2020}. Like magnetic and magneto-gravitational traps, a key advantage of Paul traps over optical trapping is that the radio frequency electric fields generated by the trap are scattering free, that is the trapped particles are not heated by the trap itself \cite{delord_2017b}.  In all cases the nanodiamonds have to be charged before loading into the Paul trap to allow trapping.

\subsection{Magnetic Field Based}\label{sec:MFB}

Magnetic traps utilise the diamagnetic nature of diamond to passively, stably trap a nanodiamond in three dimensions using a carefully-shaped magnetic potential. In this case, paramagnetic materials still cannot be stably trapped, however, diamagnetic materials can be stably trapped \cite{berry_1997}. A trap design that contains a nanodiamond particle using only magnetic fields has been demonstrated. Two rare earth magnets are machined to sharp points and brought to a vertical separation of $\approx \SI{30}{\micro\metre}$ from each other, achieving trapping frequencies of $\approx \SI{400}{\hertz}$ vertically and $\approx \SI{200}{\hertz}$ horizontally \cite{obrien_2019}. 

Magneto-gravitational traps, like magnetic traps, trap the nanodiamond using static magnetic fields, however, only in two dimensions. Trapping in the third dimension is achieved in combination with the gravitational potential. Magneto-gravitational trapping has been demonstrated for nanodiamond with trap frequencies of $\approx \SI{100}{\hertz}$ in the magnetically confined directions and $\approx \SI{10}{\hertz}$ in the gravitationally confined direction (long axis) \cite{hsu_2016}. 

\section{Conclusion}\label{sec:conc}

In conclusion, we have reviewed the literature concerning proposed experimental tests of the macroscopic limits of quantum theory, and the quantum nature of gravity, that suggest the use of a nanodiamond containing an ${\text{NV}}^{-}$ centre as a platform to generate the required macroscopic spatial superposition. Along with discussions of the proposed experiments, methods of diamond synthesis, nanodiamond fabrication, and the spin physics of the ${\text{NV}}^{-}$ centre relevant to the proposed experiments have been presented.   


\acknowledgments 
 
The authors would like to thank James March and Myungshik Kim for discussions that improved the manuscript. G. W. M. is supported by the Royal Society.  

\bibliography{spie_bib} 
\bibliographystyle{spiebib} 

\end{document}